\documentclass[aps,prb,twocolumn,longbibliography]{revtex4-1}
\usepackage{amssymb}
\usepackage{float}
\usepackage{amsfonts}
\usepackage{amsmath}
\usepackage{graphicx}
\usepackage{color}
\usepackage{latexsym}
\usepackage[colorlinks=true,citecolor=blue,linkcolor=magenta]{hyperref}

\begin{document}

\title{Engineering non-equilibrium quantum phase transitions via \\ causally gapped Hamiltonians}
\author{Masoud Mohseni}
\affiliation{Google Quantum Artificial Intelligence Lab, Venice, CA 90291}
\email{mohseni@google.com}
\author{Johan Strumpfer}\
\affiliation{Google, San Francisco, CA 94109}
\author{Marek M. Rams}
\affiliation{Institute of Physics, Jagiellonian University, \L{}ojasiewicza 11, 30-348
Krak\'ow, Poland }
\pacs{PACS number}

\begin{abstract}
We introduce a phenomenological theory for many-body control of critical
phenomena by engineering causally-induced gaps for 
quantum Hamiltonian systems. The core mechanisms are controlling information flow within and/or between
clusters that are created near a quantum critical point.
To this end, we construct inhomogeneous quantum phase transitions via designing spatio-temporal
quantum fluctuations. We show how non-equilibrium evolution
of disordered quantum systems can create new effective correlation
length scales and effective dynamical
critical exponents. In particular, we construct a class of causally-induced
non-adiabatic quantum annealing transitions for strongly disordered quantum
Ising chains leading to exponential suppression of topological defects
beyond standard Kibble-Zurek predictions. Using exact numerical techniques
for 1D quantum Hamiltonian systems, we demonstrate that our approach
exponentially outperforms adiabatic quantum computing. Using Strong-Disorder Renormalization
Group (SDRG), we demonstrate the universality of
inhomogeneous quantum critical dynamics and exhibit the reconstructions of causal
zones during SDRG flow. We derive a scaling relation for
minimal causal gaps showing they narrow more slowly than any polynomial with increasing
size of system, in contrast to stretched exponential scaling in
standard adiabatic evolution. Furthermore, we demonstrate similar
scaling behaviour for random cluster-Ising
Hamiltonians with higher order interactions. 

\end{abstract}

\maketitle

Controlling non-equilibrium dynamics of quantum many-body
systems is one of the main challenges in condensed matter physics and
quantum control. Such complex quantum systems have very rich
parameter space and unusual dynamical properties that makes them very hard
to simulate and control as they are driven through critical regions
\cite{Eisert2015}. The main difficulties arise from the fact that these systems generally
contain high degree of disorders and effectively low dimensions such that
they are not prone to exact analytical treatment or mean-field
approximations. In principle their dynamics can be mapped to the
dynamics of spin-glass systems that are driven/quenched by external
control fields and could 
experience various first and second-order quantum
phase transitions and Griffiths singularities \cite{RiegerYoung1996,vojta2006rare}. Quantum dynamics of
such complex systems, except trivial cases, would be out-of-equilibrium when
they are quenched in any finite time. However, such rich dynamical
properties could lead to novel
computational resources \cite{Nishimori1998,Farhi2001,Boixo2016,Vadim2018} provided that we obtain sufficient degree of control over their dynamics. 

\begin{figure} [h!]
\begin{center}
  \includegraphics[width=0.95 \columnwidth]{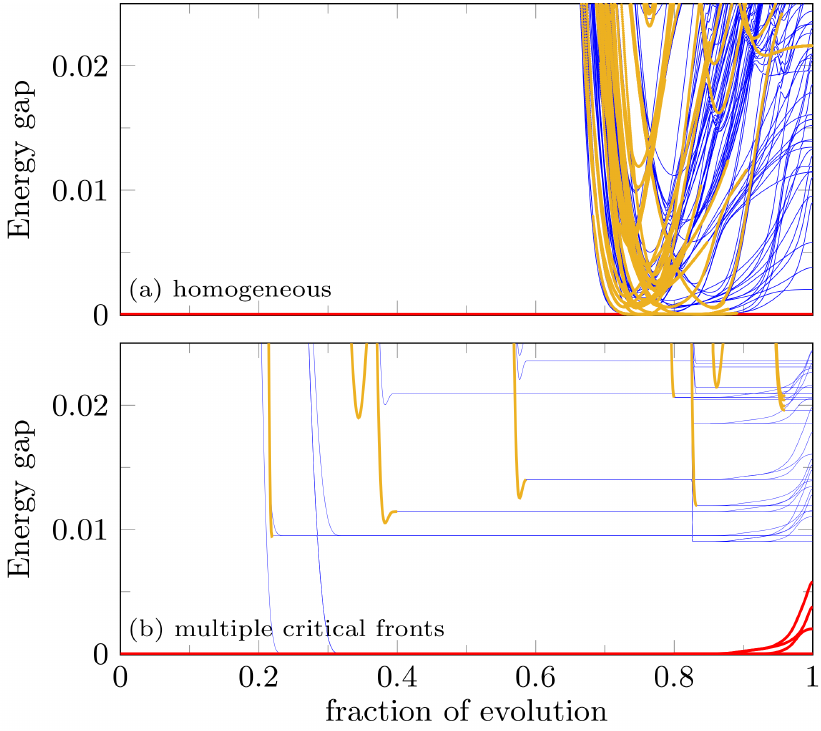}
 \end{center}
\caption{Exact numerical simulations of instantaneous eigenenergies of
  a random quantum Ising model as a function of 
evolution time for
  two distinct algorithms: (a) Homogeneous or standard
  adiabatic quantum evolution, (b) Inhomogeneous non-adiabatic quantum
  annealing. In (a) the instantaneous ground state energy is represented
  by red lines and it is set at
  $0$. Yellow marks the instantaneous excited energy states that can be reached if 
  evolution time-scale becomes comparable to the inverse of
  instantaneous gap. In
  contrast blue lines shows unaccessible excited eigenenergies due to
  vanishing instantaneous Hamiltonian state-to-state transition moments
  (see section \ref{sec:gapscaling}). In panel (b) it can be observed that the
  complexity of problem can be substantially changed via inducing
  certain causally-induced local gaps separating low-energy sector (red
  lines) from instantaneously accessible excited states (yellow lines).}
  \label{fig:multischeme}
\end{figure}

Adiabatic quantum computation (AQC) has been developed as a particular paradigm that utilize the
continuous-time dynamics of driven many-body quantum systems for solving
optimization tasks \cite{Nishimori1998,Farhi2001}. In this model, the solution of a hard
combinatorial optimization problem is encoded in the ground state of an
interacting many-body system which can be prepared
adiabatically from an initially trivial ground state, provided that time
evolution is much longer that the inverse of minimum gap square \cite{Farhi2001}.
One of the major challenges to AQC, that has been largely ignored in the
quantum computing literature, is that for many realistic problems the analog
quantum annealer will inevitably contain a significant amount of quenched
disorder smearing the corresponding quantum phase transitions for pure
systems. Thus, the required time-scale for satisfying adiabatic limit could
grow as a stretch exponential due to Griffiths singularity \cite{vojta2006rare}, even in
the absence of any first order phase transitions. The Griffiths effects have
pronounced consequences for finite-dimensional quantum systems, much stronger
than in the classical counterparts. In fact, near-term quantum processors are best 
examples of low-dimensional quantum systems due to the inherent
locality of physical interactions and geometrical constraints on the degree of
connectivity \cite{Mohseni2017}. After the embedding of a computational problem
into quantum annealers, or their digital simulations
\cite{Barends2016}, they will inevitably react to quantum fluctuations
inhomogeneously at the physical level. Consequently, near-term quantum
processors will typically experience locally inhomogeneous
and smeared first and second order phase transitions, even if we drive
them with an external field which is homogeneous in space. In
particular, annealing
schedules exhibit multiple vanishing gaps between ground state and first excited state, see Fig.~\ref{fig:multischeme}(a), leading to
exponentially long annealing time-scales. In practice, we always have a
finite annealing time-scale that would inherently violate the adiabaticity
condition, even for finite-size systems, leading to emergence of domain
walls or topological defects that emerge at a relatively wide effective quantum
critical region. This is in sharp contrast to a single, well-defined quantum
critical point for pure system, where their density of defects can be
estimated via Kibble-Zurek mechanism (KZM) in the thermodynamics limit \cite{Kibble76,*Kibble80,Zurek85,*Zurek93,*Zurek96,Dziarmaga10,Polkovnikov11,Campo14}. As of today, there is no known way to guarantee the quality of
solutions, given finite space-time physical resources, and there is no
constructive or algorithmic way to improve performance for such analog
quantum information processors within a given accuracy. These issues have lead
us to the following fundamental questions: Is it possible to engineer
quantum phase transition in disordered systems by inhomogeneous control
fields to enforce spatially-induced gaps between low energy sector and
higher energy states (see Fig.~\ref{fig:multischeme}(b))?

Here, we present a general approach for controlling quantum critical
dynamics. We introduce different classes of spatial and/or temporal
inhomogeneous protocols to drive strongly disordered quantum spin chains
across a quantum phase transition and minimize the residual energy of the
final state. This is achieved by creating governing Hamiltonian with
multiple critical fronts that can synchronize the local phase
transitions in space and time. In each local region, the number of spins
that simultaneously experience the critical dynamics is controlled by the
length scale and shape of the inhomogeneity in which the magnetic field is
modulated. Causality is introduced as the main control strategy to spatially
coordinate symmetry breaking events among neighboring regions by finding the
appropriate degree of inhomogeneities and the speed of critical fronts to
reduce the number of topological defects. We explore the conditions for an
optimal suppression of domain walls and show that we can beat the standard homogeneous KZM
prediction for the density of the topological defects for strongly disordered transverse Ising
problem in 1D. Moreover, we show that these phenomena can similarly be observed for systems
with k-local physical interactions.   We demonstrate that inhomogeneous driving can be
exponentially faster for such systems than conventional (homogenous) schemes
such as adiabatic quantum annealing. Furthermore, we
show that the universality of quantum critical phenomena holds for
inhomogeneous quantum critical dynamics even in the presence of strong
disorder. 

The outline of this paper is as follows: In section
\ref{sec:KZM}, we review causal
origins of topological defects in the context of KZM for pure and
disorder systems.  In section \ref{sec:schedules}, we describe two general classes of
inhomogeneous quantum annealing (IQA), type I and type II,
for re-constructing phase transitions and present numerical results
for strong-disorder 1D transverse Ising model. In this
context, we show that AQC can be understood as a trivial form of
either type I or type II IQA. In section \ref{sec:gapscaling}, we first provide a
phenomenological theory of the emerging local gaps and its connection to
threshold velocities for critical fronts. We then derive an
expression for distribution of local gaps as a function of inhomogeneity
slope with a logarithmic correction on the system size. We 
demonstrate universality of critical fronts shapes via strong-disorder
renormalization group (SDRG) techniques. We also discuss how the shape
of inhomogeneity is related to its penetration depth into disordered phase. A
generalization of our work for k-local Hamiltonian system is presented in
section \ref{sec:clusterH}.  A detailed treatment of our work
as a generalization of KZM and discussions on lower- and upper-bounds for
the shape of critical fronts is provided in a separate manuscript \cite{MohseniRames}.
The generalization to spin-glass systems will be presented in another
subsequent work \cite{MohseniAQC2017,Mohseni2018}.

\section{Causal origin of topological defects}
\label{sec:KZM}

We start by reviewing the Kibble-Zurek mechanism (KZM) for pure  
systems (in absence of any disorders) which has been developed as the phenomenological theory to describe the breakdown of
adiabaticity in critical systems
\cite{Kibble76,*Kibble80,Zurek85,*Zurek93,*Zurek96,Dziarmaga10,Polkovnikov11,Campo14}. The theory provides a rough estimate
for the density of topological defects that arise when a quantum or classical many-body 
system is driven through a continuous critical point at a finite rate. 
The key observation is that in the vicinity of a quantum critical point a system at the thermodynamical limit
effectively stops following the adiabatic evolution for any finite quench rate -- no matter how slow it is driven. This results in emergence of universal KZ length scale which depends on the quench rate and manifests itself, among others, in the density of topological excitations. 

The time dependent evolution of system can be expressed by a Hamiltonian as: $H(g) = g H_c + H_p$ 
where $H_c$ is controllable Hamiltonian, $g(t)$ is a control parameter with value $g_c$ at the critical
point, and $H_p$ is the Hamiltonian of interest or the ``problem Hamiltonian''.  Near a critical point  the characteristic
energy scale of the system behaves as $\bigtriangleup \sim
1/|\varepsilon|^{z\nu}$ at the thermodynamical limit. The system experiences a divergence
of the equilibrium relaxation time, $\tau =
\tau_0/|\varepsilon|^{z\nu}$, as well as a divergence
of the equilibrium correlation length, $\xi=\xi_0/|\varepsilon|^{\nu}$, where $%
\varepsilon=(g_c-g)/g_c$ is the dimensionless distance to the
critical point. The $\nu$ and $z\nu$ are the critical exponents that
characterize the universality class of the phase transition. The derivation below assumes that the exponents are well defined, i.e. they do no dependent on $\varepsilon$ and describe pure power-law dependence, and that there are no other relevant long-distance scales in the problem.

The speed of information, or the speed of second sound, is on the order of the ratio of critical length-scale to the critical time-scale
\begin{equation}  \label{speed_sound}
v_{s} \sim  \xi/\tau
= (\xi_0/\tau_0) |\varepsilon|^{\nu(z-1)} .
\end{equation}
 A causal separation near a critical point for any pair of spins could
emerge if their relative distance is much larger than length scale that
the information can propagate with the corresponding second sound velocity,
$v_{s}$, for given finite quenching time interval. Consequently, choices of broken symmetry for spins
belong to two different causal zones are not necessarily related. This is the origin of topological
defects formation. The Lieb-Robinson bound \cite{LR_bound_review}, which
characterizes the maximum speed of information in quantum many-body
systems with local interactions, provides an upper-bound for
$v_{s}$. We note that $v_{s}$ can achieve its Lieb-Robinson
upper-bound when $z=1$, such as the prototypical 1D transverse Ising model. 

Within the vicinity of $g_c$ the quenched external field can be linearized in the form $g(t)=g_c(1-t/\tau_Q)$, such that
$\varepsilon(t)=t/\tau_Q$, where $\tau_Q$ is the quench rate and
the critical point is crossed at $t=0$. The parameter regime close to the
critical point in which the system is not able to adiabatically adjust to the slowly changing external field,
and effectively, to zero-th order approximation not responding, is called frozen or impulse regime.
The freezing occurs at a particular time scale $\hat{t}$ in which the relaxation
time $\tau(t)$ becomes approximately equal to quench rate
$\varepsilon/\dot{\varepsilon}$. Thus, by setting $\tau(\hat{t})=|\varepsilon(\hat{t})/\dot{\varepsilon}(\hat{t})|$ we arrive at
\begin{equation} \label{time_KZM_pure}
\hat{t}%
=(\tau_0\tau_Q^{z\nu})^{\frac{1}{1+z\nu}}. 
\end{equation}
This equation gives the KZ time-scale relevant to describe the universal behavior of the system slowly quenched though the critical point.
The corresponding length-scale is a power-law of the quench rate as well

\begin{equation}  \label{corr_KZM_pure}
\hat{\xi}=\xi[\varepsilon(\hat{t%
})]=\xi_0(\tau_Q/\tau_0)^{\frac{\nu}{1+z\nu}}. 
\end{equation} 

This length scale can be used to estimate the size of the domains in
the broken symmetry phase. Consequently, the density of defects is
expected to vanish as $d\sim \xi(\hat{t})^{-D}$, where $D$ is the
dimensionality of system and we assume that the defects are
sufficiently robust and do not relax quickly during the subsequent
evolution.  This is the key predication of KZM. For example, in the
well-studied 
case of 1D Ising model in absence of any 
disorder we have $\nu=z=1$. The KZM prediction for the density of excitations
reads $d\sim \xi(\hat{t})^{-1} \sim \tau_Q^{-1/2}$ in that case \cite{Dorner2005,Dziarmaga2005,Polkovnikov2005}, which can indeed be verified analytically\cite{Dziarmaga2005}. The above argument was later developed into full dynamical scaling hypothesis, which allows to obtain similar power-laws for other observables of interest \cite{Deng_EPL_2008,DeGrandi2010,Kolodrubetz2012,Chandran2012a,*Chandran2012b,Francuz2016}.

Understanding causal effects in disordered systems near a critical point and any attempt for estimation of
density of defects requires careful analysis and not much is known outside of specific cases. Experimentally, quenches from the superfluid to the Bose glass were reported \cite{DeMarco_BoseGlass_2016},
with the resulting residual energies vanishing very slowly with the increasing quench rate. Full theoretical understanding is still missing in this case.
Theoretical investigations are mostly limited to the class of systems with the critical point in the universality class of so-called infinite-disorder fixed point.
Here, we are interested in systems belonging to this class. We first consider the prototypical example of a random transverse Ising Hamiltonian for a chain of $N$ spins,  
$\hat H = -\sum_{n=1}^N g(n) \sigma^x_{n} -\sum_{n=1}^{N-1} J_{n,n+1} \sigma^z_n \sigma^z_{n+1}$,
with quenched (fixed) disorder in the nearest-neighbors couplings $J_{n,n+1}$. In this article we assume that they are drawn from the flat distribution over interval $[-1,1]$. 
The unit of time is set by $\hbar=1$.  Using strong-disorder renormalisation group (SDRG) techniques, the equilibrium properties of this model were first
evaluated by Daniel Fisher \cite{Fisher1995,*Fisher1992a}.  For a homogenous or uniform
transverse field in the model, the distribution of disorders induces a critical
point that can be evaluated by relation $g_c =  \exp(\overline{\log(\vert
 J_{n,n+1} \vert)})$. For uniform distribution of $J_{n,n+1} \in
[-1,1]$ this yields a critical value of $g_c =e^{-1} \simeq 0.367879$.
It should be pointed out that the critical point for similar systems in
two dimensions \cite{Rieger2007} and in presence of dissipation \cite{Vojta2009} are also known to belong to this
universality class. We use numerical SDRG 
to demonstrate universality of our
non-equilibrium protocols in the section \ref{sec:gapscaling}. We also generalize our results to Ising model with certain k-local interactions in the section~\ref{sec:clusterH}. 

The presence of disorder, changes the
universality class of the critical point of the Ising model from $\nu=z=1$ to $\nu=2$ and $z
\to \infty$, and thus quantitatively and qualitatively modifies the
dependence of correlation length and density of
defects on the quench time-scale. 
Most importantly, using SDRG techniques, it
was evaluated that as the system approaches the
critical point the gap of random Ising model scales as $\bigtriangleup[\varepsilon] \simeq
|\varepsilon|^{1/|\varepsilon|}$ [\onlinecite{Fisher1995,*Fisher1992a}], and consequently
the critical exponent  $z\nu= 1/|\varepsilon| + O(1)$ diverges as $\varepsilon \to 0$. 
For that reason the KZM derivation described earlier has to be modified to take this into account \cite{Dziarmaga2006,Caneva2007}.
The characteristic time-scale  $\hat{t}$ follows from the condition
$|\dot{\varepsilon}(\hat{t})/\varepsilon(\hat{t})| = 1/  (\tau_Q |\varepsilon(\hat{t})|)
\approx \kappa |\varepsilon
(\hat{t})|^{1/|\varepsilon(\hat{t})|} $, where $\kappa$ is a constant
factor on the order of one. The above relation can be solved in the limit
of infinitely long annealing time, $\ln(\tau_Q) \gg 1$, yielding
\cite{Dziarmaga2006}

\begin{equation}  \label{corr_KZM_disorder}
\hat{\xi} \sim
\frac{\ln^2(\tau_Q/\kappa)}{\ln^2[\ln(\tau_Q/\kappa)]}. 
\end{equation}

The density of defects is then suppressed logarithmically with quench time $d\sim1/\ln^2\tau_Q$, which is quadratically faster than
simulated annealing, where defects scale as $d\sim1/\ln\tau_Q$ \cite{Suzuki2009,Santoro2015}. 
The existence of these logarithmic scaling laws implies that one has
to run exponentially long annealing times to reduce the residual energy
of the final state. However, as we will show in the next section one can
recover a polynomial scaling by driving the system with a spatially inhomogeneous
transverse field. 

\section{Causal control of topological defects with multiple
critical fronts}
\label{sec:schedules}

From carefully studying defect formation under homogeneous drive fields, one can see how a new way
of suppressing or controlling topological defects can emerge by being
aware of causal separation of 
subsystems due to the extremely small
values of velocity for information propagation near a critical point
according to Eq.~\eqref{speed_sound}.  In other words, one can try causal
synchronization of the local phase transitions by 
inhomogeneous driving fields, as far as the 
critical front do not move faster than a threshold velocity corresponding to the speed of information,
see \cite{Dziarmaga2010a, Dziarmaga2010b, Rams2016,shondi_critical_2017} for a quantum case and \cite{Kibble1997,Dziarmaga1999,Zurek2009,ions10,DRP11,DKZ13} for classical counterpart. Note that this is fundamentally
different than the standard annealing paradigm which is guided by the inverse of a \textit{global gap} of a quantum
Hamiltonian system which provides an upper-bound for relaxation time scales
according to the adiabatic theorem. In other words, adiabaticity provides a
sufficient condition for annealing time and it is not necessary to get
low-energy states or even the ground state of disordered Ising systems.  

Here we provide a phenomenological description of causally-induced
non-equilibrium quantum phase transitions. Specifically, we develop an
algorithmic quantum annealing approach to create a
causal sequence of locally gapped Hamiltonians. We
note that for strongly disordered systems in low dimensions there is a quantum
Griffiths region that is spread in the disordered and ordered phases, i.e. on both
sides of a critical point \cite{Fisher1995,*Fisher1992a}. Within the Griffiths
region the system undergoes effective local phase transitions
that are space-time separated in nature even if the control fields are
homogeneous. The key observation is that one can create situations in
which the choices of symmetry-breaking
events in a local neighborhood that have already experienced phase
transitions earlier could influence the symmetry breaking events
elsewhere, provided that the control fields have certain
inhomogeneous spatiotemporal structures. These
symmetry breaking events are perceived by the rest of the system,
which is still in a disordered phase, as effective boundary conditions influencing their local fields.

In order to develop an algorithmic quantum annealer, here we construct a
general class of inhomogeneous quantum annealing schedules.  They are a function of
a fixed total quench time or annealing time $T \sim \tau _{Q}$, proportional to the annealing rate of the homogeneous quench $\tau_Q$ introduced in the previous Section. The
performance of the algorithms are evaluated by computing the 
precision $\epsilon_Q $ of approximating the ground state. Here we mostly
focus on the random instances of strongly-disorder spin chains, nevertheless
our construction is general and can be applied to higher-dimensional
systems \cite{Mohseni2018}. There are two main reasons for such a choice. First, for 1D case we can
simulate their dynamics exactly using a mapping to free-fermionic system, as e.g. in Ref.~\onlinecite{Dziarmaga2006,Caneva2007,Rams2016}. Also the critical behavior of such systems when driven via homogeneous external fields have been studied extensively, thus the new non-equilibrium physics of such systems when driven inhomogeneously can be better benchmarked and appreciated. The overall Hamiltonian for a system of $N$ spins under a inhomogeneous driver field can be written as:
\begin{equation}\label{H_total}
H(t)=-\sum\limits_{n=1}^{N}g(n,t)\sigma
_{n}^{x}-\sum\limits_{<n,m>}^{N}J_{nm}\sigma _{n}^{z}\sigma _{m}^{z},
\end{equation}%
and the quality of an output state is characterized by a normalized residual energy as:
$\epsilon_Q =Q/N$, with $Q =  \left\langle \psi (\tau _{Q})\right\vert
H_{p}\left\vert \psi (\tau _{Q})\right\rangle -\left\langle \psi
_{gs}\right\vert H_{p}\left\vert \psi _{gs}\right\rangle$, where $\left\vert \psi (\tau _{Q})\right\rangle $ is the quantum state of
the system at the final annealing time. $\left\vert \psi _{gs}\right\rangle $
is the ground state of the classical time-independent Hamiltonian, or the
problem Hamiltonian, $H_{p}=-\sum\limits_{<n,m>}^{N}J_{nm}\sigma
_{n}^{z}\sigma _{m}^{z},$ with eigenvalue $E_{gs}=\left\langle \psi
_{gs}\right\vert H_{p}\left\vert \psi _{gs}\right\rangle $. We note that
for pure systems, where $J_{nm}=J$, the normalized residual energy can be
related to Kibble-Zurek correlation length $\hat \xi $ by $\epsilon_Q \sim
\left\vert J\right\vert \hat \xi ^{-D}$ where $\hat  \xi ^{-D}$ is the density of
topological defects and $D$ is the dimension of system.

Here we assume that the inhomogeneous drive field is a transverse field that
can be locally modulated for every individual spin. This Hamiltonian can be
realized with the near-term quantum annealing technologies currently being
developed at the D-Wave Quantum Computing Systems and Google Quantum AI Lab.

As we describe in the next section, for any given instance of disorders $%
\left\{ J_{nm}\right\} $ as we drive the system toward the quantum
critical point, the system responds to quantum fluctuations within $M$ distinct ``clusters", which are related to the emergence
of rare local regions within the Griffiths phase. As we will show, the number and locations
of clusters can be estimated via a simple preprocessing step that grows linearly
with the size of the chain for 1D system. The
generalization to higher dimensional system is presented in Ref. \cite{Mohseni2018}.

In each cluster we drive the many-body system by a transverse Ising
Hamiltonian with some local structure. Thus, we drive these $M$ clusters
simultaneously into some space-time separated inhomogeneous transitions,
\begin{equation}\label{g_nt}
g(n,t)=\sum\limits_{l=0} h_{l}(n)\lambda_{l}(t)+\sum\limits_{k=1}^{M}\omega
_{k}g_{k}\left(||n-n_{k}||-v_{k}(n,t)t\right),
\end{equation}%
where each $h_{l}$ is the time-independent global magnetic field which has a
spatial structure and each $\lambda_{l}(t)$ is
spatially uniform but it can generates nonlinear dependence to
time. The terms $\sum\limits_{k=1}^{M}\omega
_{k}g_{k}\left(||n-n_{k}||-v_{k}(n,t)t\right)$ characterize various spatiotemporal
dependencies of traveling quantum critical fronts, where $||n-n_{k}||$ denotes a distance measure of node $n$
from some center node $n_{k}$ per cluster where we trigger the quantum
fluctuations. The center of these spatiotemporal inhomogeneities can
be shifted linearly in time by $v_{k}(n,t)t$ with a spatiotemporal
dependence for each $k$ cluster. However, for simplicity, for rest of
this work we concentrate on a constant critical front motion
for each cluster; that is $v_{k}(n,t) = v_k$.

In the following section, we define the shape and two different kind of
velocities for critical fronts and provide two simple examples of type
I and II annealing, namely periodic inhomogeneous annealing, and
mutliple-critical-fronts inhomogeneity.

\subsection{Shape and velocity of critical fronts}

The inhomogeneity slope and its horizontal and vertical velocities of
inhomogeneity can be characterized by a set of hyper-parameters $\left\{
\alpha ,v^{h},v^{v}\right\}$ corresponding, respectively, to local slope of the instantaneous field in space and it's spatial (horizontal) and temporal (vertical) velocities,
that are defined by derivatives of $%
g(n,t)$ and $n(g_{fix},t)$ as:

\begin{eqnarray} \label{alpha_v_defs}
v^{v}(n,t) &=&-\partial g(n,t)/\partial t, \\
v_{k}^{h}(n,t) &=&\partial n(g_{fix},t)/\partial t, \\
\alpha (n,t) &=&\partial g(n,t)/\partial n.
\end{eqnarray}
Thus, we can derive closed form expressions over these hyperparameters 
for different annealing schedules. To appreciate the generality of
the shape of $g(n,t),$ we consider two concrete and qualitatively
distinct classes for inhomogeneous quantum annealing with respect to possible temporal and/or spatial inhomogeneities.

\begin{figure} [t]
\begin{center}
  \includegraphics[width=1 \columnwidth]{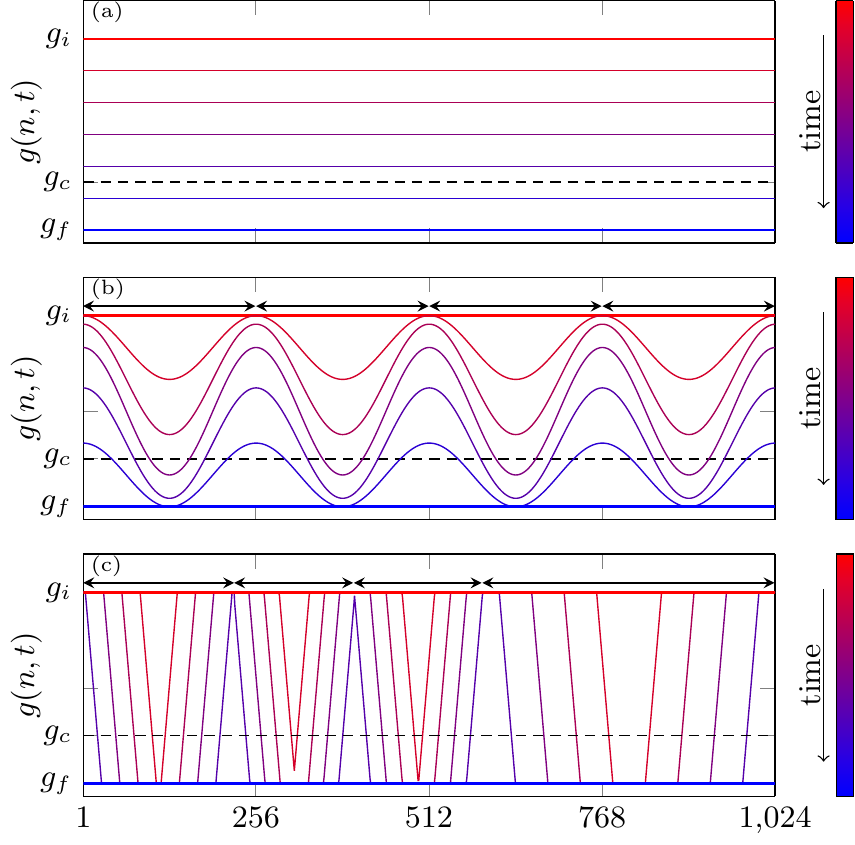}
\end{center}
  \caption{{Illustration of three different annealing protocols}. Different
    lines indicate different snapshots of time-dependent magnetic
    field as it has been driven from the initial to the final value across a
    quantum critical point. { Top panel} shows the standard or
    homogeneous transverse field which can be also considered as a trivial
 case of both type I and II annealing. { Middle panel} illustrates
 the periodic driving of type I inhomogeneous annealing as in Eq.~\eqref{eq:g_sinwave} with 4 clusters and amplitude $a=1$. 
 { Lower panel} shows a prototypical example of type
 II inhomogeneous quantum annealing with $4$ clusters. The borders
 between clusters is tuned to coincide with the weakest values of
 $J_{n,n+1}$ for a given realization of disorder as a result of a
 simple preprocessing procedure. Here we have chosen a constant
 inhomogeneity slope of $\alpha=1/8$ for each cluster. 
 The dashed lines
 in each figure show the critical value of the transverse field.}  
  \label{fig:schedules} 
\end{figure}

\subsection{ Type I inhomogeneous quantum annealing: space and time separated 
inhomogeneity}

In this class, we consider a general form of independent or separated
space and time quantum fluctuations to drive the annealing dynamics
\begin{equation}\label{IQA_typeI}
g(n,t)=\sum_{l=0} h_{l}(n)\lambda_{l}(t).
\end{equation}
An example of this class will be a periodic spatial inhomogeneity (standing wave) combined
with spatially-independent time-local inhomogeneity as:
\begin{equation*}
g(n,t)= h_0 \lambda_0(t) + \lambda_{1}(t) \sum\limits_{k=-\infty }^{\infty }a_{k}e^{i\pi kn},
\end{equation*}%
where each term in the spatial contribution in the second term corresponds to an
estimated cluster size. We provide a simple illustration of these periodic spatial inhomogeneities in the next section. 
We note that KZM -- in the context of pure systems -- was also extended to quenches that are homogeneous in space, but nonlinear (inhomogeneous) in time\cite{Mondal2008,*Mondal2009,*Polkovnikov2008,Chandran2012a,*Chandran2012b}. Such inhomogeneity adjusts the quench rate to the distance from the critical. Consequently it allows to reduce the number of generated defects.

\subsection{Type II inhomogeneous quantum annealing: Spatiotemporal inhomogeneities}

In this class, we build a sufficiently general example by creating a
mutliple-critical-fronts annealing schedule in $M$ clusters where
critical fronts in each cluster are moving 
with the speed $%
v_{k}(n)$ and each are governed by a separate activation function $\tanh
\left[\theta _{k}(||n-n_{k}||-v_{k}(n)t)\right]:$

\begin{equation}\label{IQA_typeII}
g(n,t)=g_{c}\left\{1+\sum\limits_{k=1}^{M}\omega _{k}\tanh \left[\theta
_{k}(||n-n_{k}||-v_{k}(n)t)\right]\right\}
\end{equation}%
where $g_{c}$ is the critical value of transverse field. We note that
there is no particular significance for our choice of activation
function here. As an important special class of the above driver
field, we linearize the activation function in each cluster near quantum
critical point, that is $\tanh [\theta _{k}(n-v_{k}(n)t)]\simeq $ $\theta
_{k}(n-v_{k}(n)t)$, then for each cluster we get:
\begin{equation}\label{IQA_typeII_linear}
g_{k}(n,t)=g_{c}\{1+\theta _{k}(n-v_{k}(n)t)]\}.
\end{equation}

In the first example of this type, we consider an inhomogeneity with
constant $v_{k}$ for each cluster of the form $g_{k}(n,t)=g_{c}\{1+\tanh
[\theta _{k}(n-v_{k}t)]\}$ which yields $n=\tanh
^{-1}(g_{k-fix}/g_{c}-1)/\theta _{k}+v_{k}t.$
\begin{eqnarray*}
v^{v}(n,t) &=&g_{c}[1-\tanh ^{2}[\theta _{k}(n-v_{k}t)]]\theta
_{k}v_{k}=\alpha _{k}(n,t)v_{k}, \\
v_{k}^{h}(n,t) &=&v_{k}, \\
\alpha _{k}(n,t) &=&g_{c}[1-\tanh ^{2}[\theta _{k}(n-v_{k}t)]]\theta _{k}.
\end{eqnarray*}

In the next example of this type, we consider linear approximation of activation
function near critical point which yields $g_{k}(n,t)=g_{c}\{1+\theta
_{k}(n-v_{k}t)]\}$ and $n=(g_{k-fix}/g_{c}-1)/\theta _{k}+v_{k}t.$ Thus we
have: $v^{v} =g_{c}\theta _{k}v_{k}=\alpha _{k}v_{k}$, 
$v_{k}^{h} =v_{k}$, and $\alpha _{k} =g_{c}\theta _{k}$.

\begin{figure} [t]
\begin{center}
  \includegraphics[width=0.95 \columnwidth]{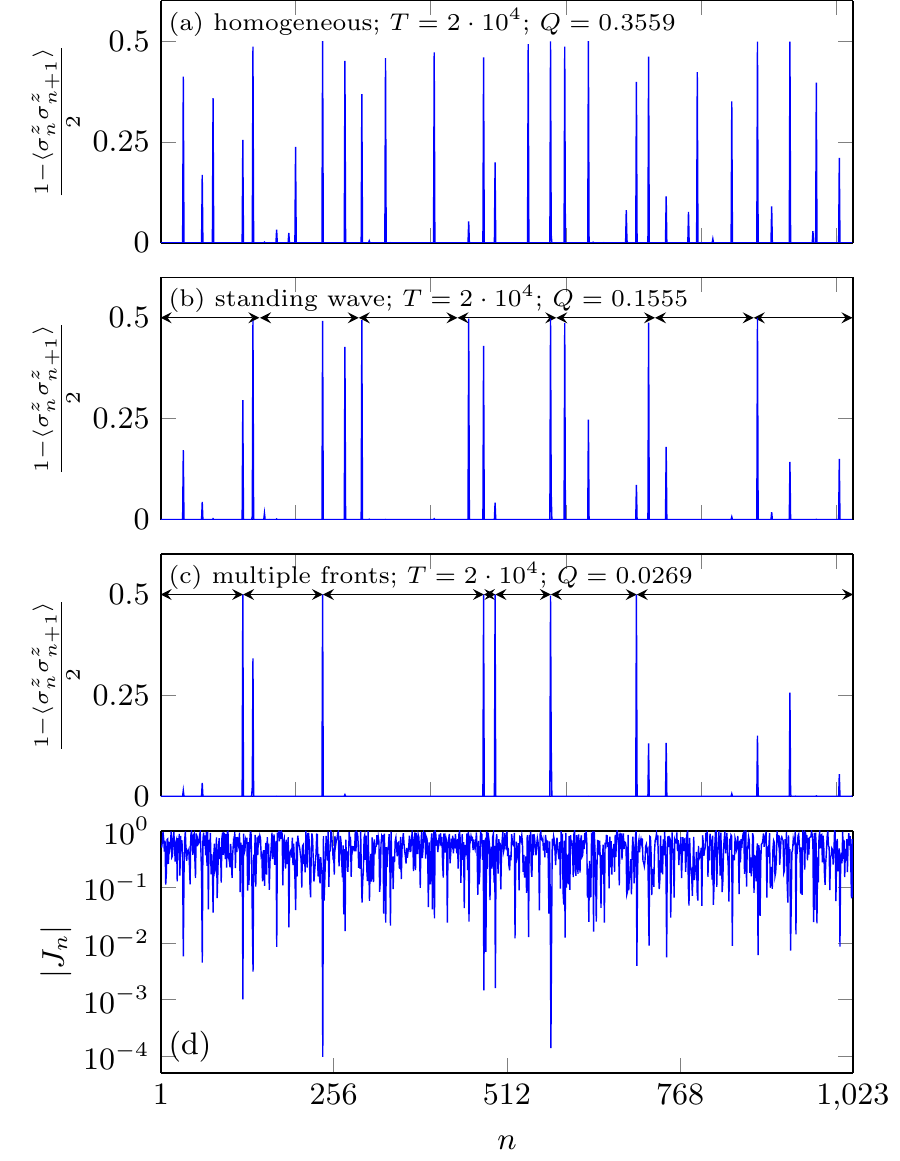}
 \end{center}
  \caption{{ Comparing distribution of topological defects for different
      protocols applied to a single instance of disorder Ising
      model at $T=2\cdot10^4$}. This instance is chosen to be typical in a
    sense that residual energies are close to respective median values
    for either strategies. { (a)}: Homogeneous strategy where density of defects is
    expected to diminish as $\log^{-2} T$. { (b)}: Defects
    distribution for type I
    inhomogeneous protocol from Eq.~{\eqref{eq:g_sinwave}} with 7
    clusters of the same size.  { (c)} Defects distribution for a
    type II annealing schedule where the clusters
    are formed is such a way that the weakest links resides at the
    borders. Inhomogeneous strategy essentially brushes the defects and
    place them at the borders between clusters where the energy penalties are
    minimized. We use $\alpha=1/8$. 
    { (d)}. For reference we show
    the values of random couplings for the corresponding realization
    -- notice logarithmic scale which
 is emphasizing the position of the weakest links. }
  \label{fig:defects}
\end{figure}

\subsection{Standard AQC: Absence of any inhomogeneity}

We note that the standard or homogeneous quantum annealing schedule which
has been extensively studied and numerically benchmarked for almost two
decades can be considered as the extreme limit of
either type-I or type-II of IQA. In the former case we have one spatially
uniform transverse field, $g_{0}(n)=\mathrm{const.}$,
and linear velocity $\lambda_{0}(t) =(1-t/\tau _{Q})$, where $\tau _{Q}$ is the overall
annealing time-scale.  Thus we have the
familiar form of homogeneous transverse field, which is linear in
time, as: $g(n,t)=g_{0}(1-t/\tau _{Q})$. In order to see AQC as a extreme limit of type-II IQA, we must note
that the homogeneous transfer field can be considered as a single critical
front with a trivial flat shape with infinite velocity; that is
$\theta \rightarrow 0$ and $v(n) \rightarrow \infty$, while
$\theta v(n)$, which is basically the vertical
velocity, will be equivalent to the inverse of annealing time
and thus will be finite. The hyperparameters for homogenous
annealing respectively become  $v^{v}(n,t) =1/\tau _{Q}$, 
$v_{k}^{h}(n,t) =\infty$, and $\alpha (n,t) =0$.

\subsection{Exponential suppression of defects} 

To illustrate the power of multi-front critical control, we numerically investigate
two concrete forms of type I and II inhomogeneous annealing 
as described above and compare their performance against standard QA. All
the simulations in this section are done using the Jordan-Wigner transformation
that maps the Hamiltonian in Eq.~\eqref{H_total} onto the system of
free fermions where it can be solved numerically in a standard way. For
details of these techniques, we refer the readers to the Appendix B of
Ref.~\onlinecite{Rams2016}. 
For our numerical analysis here the cluster formation that we invoke is simple and has linear
scaling with the system size for Ising chains. In the next section we
use SDRG to examine construction and scaling of causal gaps.

Examples of the type I and II annealing schedules for one random
instance of Ising chain are
given in Fig.~\ref{fig:schedules}, in which different snap-shots of
time-dependent transverse fields are plotted along the chain. In
Fig.~\ref{fig:schedules}(a), we illustrate the trivial/homogeneous
schedule. In Fig.~\ref{fig:schedules}(b) we explore the
effects of periodic critical fronts by constructing the schedule   
\begin{equation}
\label{eq:g_sinwave}
g(n,t) = g_i(1 - t/T) + a \cos\left(kn \right) \sin(\pi t /T),
\end{equation}
where $T$ is the total evolution time.
Finally, in Fig.~\ref{fig:schedules}(c) we illustrate an example of
multiple-critical-fronts strategy. 

In the latter, in order to decide the position of the cluster, we employ a simple pre-processing.
It is based on observation, that for strictly 1D geometry, the
instantaneous local gap may be set 
by a single, very weak link.  We discuss it in more detailed in Sec. \ref{sec:gapscaling}. Such weak link sets the local time-scale needed for adiabatic transition dividing the chain into two weakly interacting parts. We want to place the borders between the clusters at such links, as (i) they would require the longest time to align according to the weak coupling and (ii) the energy penalty for placing the defects there is the smallest. To that end, we look for the largest cluster -- starting at the end of the chain or at the end of the previous cluster --
where $\min_{cluster_k}(J_{n,n+1})\cdot \kappa > v_k$. Here, $v_k$
would be the velocity of the front for this ($k$-th) cluster, and
$\kappa$ is a parameter fixing the exact value of the threshold (in
practice $\kappa \approx 2$). This condition would allow for adiabatic
transition as if the energy gap were set by single links only.
If the condition is not satisfy the considered candidate for cluster is cut at its weakest link, creating a new smaller cluster where we check the condition again. The procedure is repeated until full chain is divided into clusters.
As a result, for a total fixed annealing time, all velocity $v_{k}$ are cluster
dependent; allowing optimization of the computational parameters over
the available time. 
For a cluster of size $L_{k}$ sites the (vertical)
velocity is given by  $v_{k}^{v}= (|g_f - g_i| + \alpha L_{k} /2)/T$ for a given
fixed total annealing time $T$. Each cluster
is driven separately, and the inhomogeneous front is brushing from the middle of each cluster to both ends simultaneously.
For each cluster spanning spins $n = 1,2 \ldots L$ (counting from the beginning of the cluster) the magnetic field is
constructed as:
\begin{equation} \label{gn_inhomo_multi}
g(n,t) = \left\{ 
\begin{array}{ll}
g_i, & |n- \frac{L}{2} | - t v > \frac{g_i-g_f}{2 \alpha },   \\
\frac{g_i+g_f}{2} + \alpha ( |n- \frac{L}{2} | - t v) , & \left| |n- \frac{L}{2} | - t v \right|\le \frac{g_i-g_f}{2 \alpha},    \\
g_f  & |n- \frac{L}{2} | - t v < \frac{g_f-g_i}{2 \alpha },
\end{array} \right.
\end{equation}

\begin{figure} [t]
\begin{center}
  \includegraphics[width=0.95 \columnwidth]{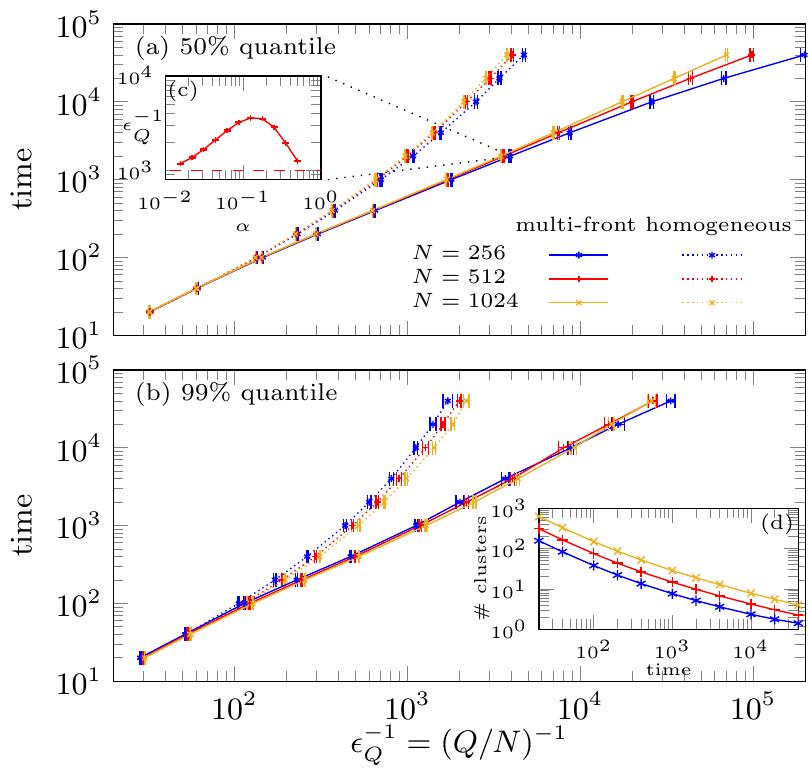}
 \end{center}
  \caption{{ Scaling of the total annealing time as a function of inverse
      residual energy for homogeneous control (dashed lines) versus
    multiple critical fronts of phase transitions (solid lines)}.
  Various colors/symbols indicate different system sizes: blue asterisks  for
    $N=256$, red pluses for $N=512$, and yellow crosses denoting $N=1024$. Panel (a)
    shows median ($50\%$ quantile) and panel (b) shows $99\%$ quantile. Statistics is build from 1000 instances of
    disorders. An exponential improvement in annealing time to obtain
    a fixed residual energy is observed from $\epsilon_Q \sim (\log
T)^{-3.8} $ for adiabatic quantum annealing to $\epsilon_Q \sim
T^{-1.03}$ for inhomogeneous non-adiabatic annealing. The slope of
all multi-front
schedules are chosen to be at optimal value of $\alpha
= 1/8$, see panel (c). Panel (d) shows the mean
number of clusters for type II annealing for different time-scales resulting from preprocessing procedure.}.
  \label{fig:timescaling}
\end{figure}

The probability distribution of topological defects for these protocols are
presented in Fig.~\ref{fig:defects}. The defect density is
evaluated for a strongly disordered instance of a spin chain consist of up to
about 1000 qubits with $J_{n,n+1}$ sampled randomly from $[-1,1].$
Here, the inhomogeneous
annealing can be regarded as a many-body quantum control strategy
which can significantly reduce the number of topological defects by
synchronizing the symmetry breaking events and can brush the reminder of
defects into the weakest $J_{ij}$ where they act as defect
sinks. In other words, not only do these non-adiabatic paths suppress the
emergence of domain walls, but also minimize the energy cost per defect by several order of magnitudes.

The scaling of annealing or quenching time as a function of inverse residual energy
is plotted in Fig.~\ref{fig:timescaling} for 1000 random instances of 1D Ising
chains ranging from 256 spins to 1024 spins. It can be observed that
the annealing time is improved exponentially over standard AQC scheme
for typical, $50\%$ quantile, as well as harder instances, $99\%$
quantile (each corresponds to the 
residual energy in which $x\%$ instances
have smaller values).  In the homogeneous protocol we have $\epsilon_Q \sim (\log
T)^{-\gamma} $. We obtain $\gamma \approx 3.8\pm0.4$. It can be compared with $\gamma \approx 3.4$ which has been obtained by  from
Caneva {\it at. al.} \cite{Caneva2007} from smaller values of quench times $T$ and for slightly different protocol, where the magnetic field was also disordered.
What is important here is that $\gamma$ is larger than $2$, i.e. the value of exponent governing the scaling of defect density. This results from defects being more likely to appear on the links with smallest $|J_{n,n+1}|$. For uniform distribution in $[-1,1]$, $\gamma=4$ would correspond to defects appearing only at the weakest links.

 On the other hand, in type II protocol we
observe $\epsilon_Q \sim T^{-1.03}$ (fit for $N=1024$, $T>100$
and $50\%$ quantile).  Multi-front protocol have been constructed by fixed $\alpha
= 1/8$, where optimality of such choice is shown in panel (c) for
$T=1000$. Similar plots for other time-scales (not shown) suggest
that in this system this value is optimal independently if $T \gg
1$. Panel (d) shows the mean number of clusters in type II annealing
for different time-scales. For large times it should scale as $\sim
{N}/{T^{0.5}}$ which results from the preprocessing procedure. This follows from the expected size of clusters which can be solved in given $T$ for $J_{n,n+1}$ drawn from a uniform distribution in $[-1,1]$.

\section{Scaling relations for causal gaps}
\label{sec:gapscaling}

Here, we introduce the notation of causally gapped Hamiltonians that
are created by time-dependent multiple critical fronts introduced in
the previous section. In particular, we derive a scaling relation for
the distributions of minimum causal gaps as the function of system
size and inhomogeneity slope. In this work, we define the
\textit{causal gaps} as the inhomogeneously-induced instantaneous energy gap which becomes
relavant when the critical front is driven below a threhold velocity allowing for
information of symmetry-breaking events to propagates. 
The core idea is that there is an
effective threshold velocity $v_t^{k}$ that determines the
suppression of topological defects formation within each cluster in
disordered systems. If the velocity of the front in each cluster is
much larger than this threshold speed, $v_k \gg v_t^{k}$ for all $k$, then the
effect of the spatial variations of the control field become irrelevant
in the sense that we will recover the standard critical dynamics created by
homogeneous driving, which can be understood
by the standard KZM. However, 
when we drive each critical front such that $v_k \ll v_t^{k}$, the length scale
and shape of the critical front becomes highly relevant allowing to suppress creation of the topological defects in each cluster.
The shape of the critical front determines the number of spins that simultaneously experience criticality creating an effective (finite-size) energy gap.
Otherwise, the homogeneous system would be gapless at the critical
point in the thermodynamical limit.

\subsection{Causal gap for pure systems}

In the absence of disorder and with sufficiently smooth critical front, that is $\alpha \ll 1$, one can invoke a variant of KZM to estimate when the inhomogeneity of the driving front is relevant.
This question can be regarded from two perspectives
\cite{Dziarmaga2010a,Dziarmaga2010b}.

Firstly, starting from the limit of homogeneous driving, we note that the relevant speed of
information at the critical front can be expressed as $\hat{v} \sim
\hat{\xi}/\hat{t}$. Here, $\hat{\xi}$ and $\hat{t}$ are the effective
length scale and time scale that the system experience according to
KZM given by the relations \eqref{time_KZM_pure} and \eqref{corr_KZM_pure}, respectively. This yields 
$\hat{v} = \tau _{Q}^{\nu(1-z)/(1+z\nu)}$. Next we consider the control parameter $\varepsilon(n,t)$ to be position dependent. We linearize the relation at the critical front for fixed position $n_{fixed}$ as 
\begin{equation}
\varepsilon(n,t) = \alpha(n_{fixed}-vt)=-\alpha v t +\mathrm{const}.
\label{eq:epsilon_linearized}
\end{equation}
 This gives the local annealing rate $\tau_Q(n)=1/(\alpha v)$. 
Causality implies that if $v  \gg \hat v$, then the choice of symmetry breaking which happened earlier along the chain cannot influence what is happening later and we recover the independent defect formation assumed in the standard KZM. The self-consistency condition allows to express such threshold velocity as a function of our main control parameter $\alpha$. This is obtained by inserting the above annealing rate into
the expression for $\hat{v}$, which leads to
\begin{equation} \label{vc_KZ}
\hat{v_t} \sim \alpha^{\nu(z-1)/(1+\nu)}.
\end{equation}

\begin{figure} [t]

\begin{center}
  \includegraphics[width=0.95 \columnwidth]{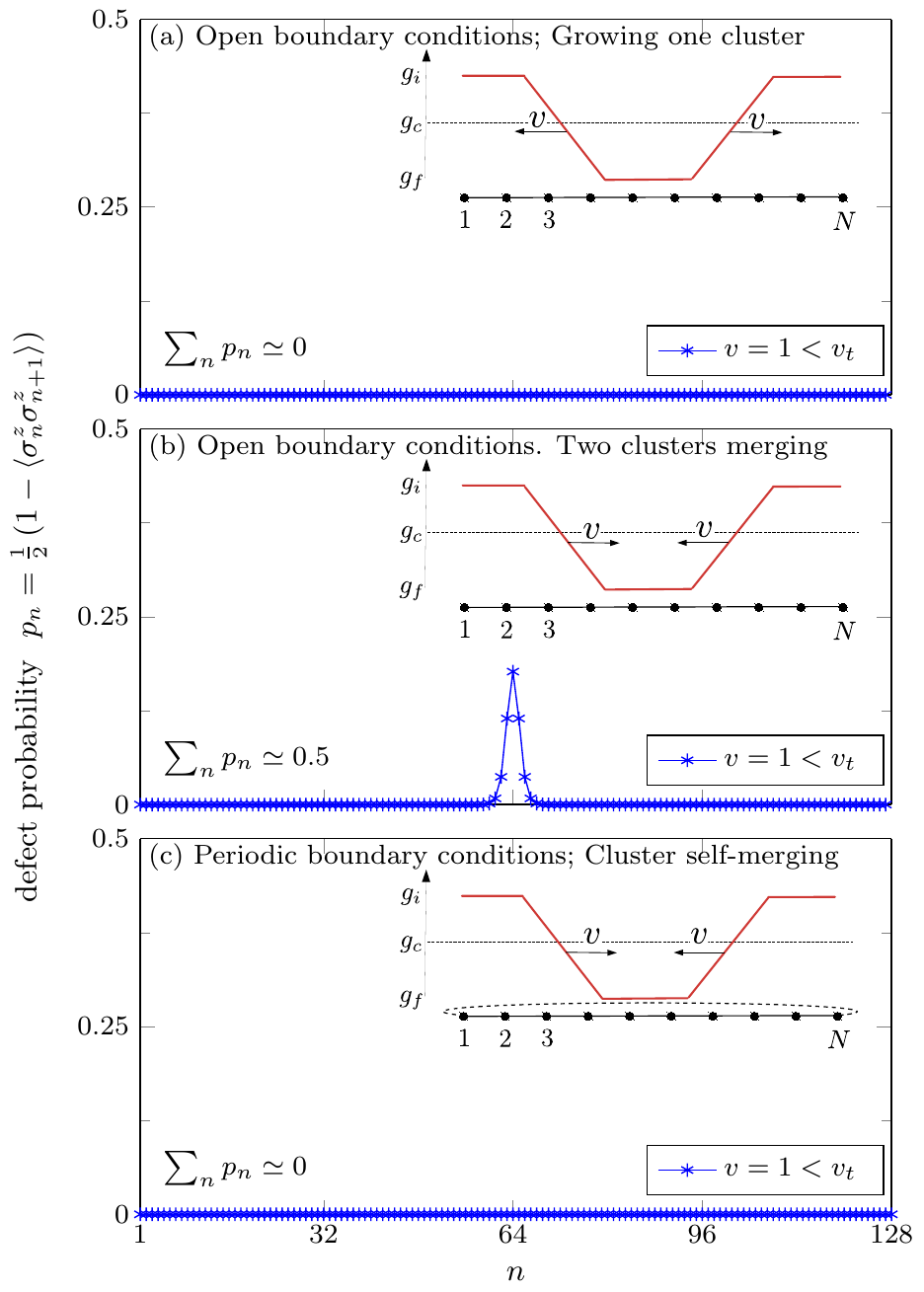}
 \end{center}
  \caption{{ Defects formation as the clusters are merging:} We consider a pure Ising model for illustrative purpose. { In (a)} we consider 1D chain with open boundary conditions and a single cluster growing from the center of the chain. The critical fronts are moving with vertical velocity below the threshold and, as a result, we obtain a system without defects after the quench.
{ In (b)} we consider the setup with two clusters and two fronts merging in the middle of the chain. Again, the fronts velocity is below the threshold and there are no defects inside clusters. However, as the clusters were independent, defect can be created as they merge, reflecting the possibility that two independent clusters might break the symmetry differently. The probability of having such defect is equal $1/2$ as can indeed be seen from the numerics.
{ In (c)} we consider the system with periodic boundary
conditions. Effectively, in this setup we have one cluster which is
self-merging at the end of the quench. Importantly, such process does
not lead to the creation of a defect -- in contrast to the situation
in panel (b).}
   \label{fig:causalcartoon}
\end{figure}

Alternatively, we could look at the instantaneous Hamiltonian resulting from inhomogeneous front in Eq.~\eqref{eq:epsilon_linearized}. We focus here on the instantaneous ground state of such system, which is interpolating -- in space -- between order and
disorder phases. They are spatially separated by an effective critical regime which size, called a \textit{penetration depth}, can be estimated as
\begin{equation}
\hat \xi_i \sim \alpha ^{-\frac{\nu}{\nu+1}}.
 \label{eq:IKZM_xi}
\end{equation}
It follows from a variant of KZM argument,  so called KZM in space \cite{Turban2007,ZD08,Turban2009,Damski2017}, where one asks about characteristic distance from $n_{fixed}$ up to which the system is able to locally adjust to $\varepsilon(n)$ changing in space as if it were locally homogeneous with local correlation length determined by $\varepsilon(n)$. Apart from the Ising model \cite{Turban2007,ZD08,Turban2009,Damski2017}, the interplay between inhomogeneous external field and criticality was studied in the context of spin$-1$ Bose-Einstein \cite{Damski2009}, the XY model \cite{Vicari2010}, and the XXZ model \cite{Dutta2015}.

The finite size of the effective critical region in Eq.~\eqref{eq:IKZM_xi} allows to estimate the gap of the instantaneous Hamiltonian as
\begin{equation} 
\hat \Delta_i \sim \alpha ^{\frac{z \nu}{\nu+1}}.
\label{eq:iKZM_Gap}
\end{equation}
By combining those characteristic scales we obtain a threshold velocity $v_t$ which is again given by Eq.~\eqref{vc_KZ}. The meaning of this threshold velocity is however different here. Namely, we can expect that if
the velocity $v$ in Eq.~\eqref{eq:epsilon_linearized} is $v \ll v_t$,
then the system would be able to follow its instantaneous ground
state.

\begin{figure*} [!t]
\begin{center}
  \includegraphics[width =\textwidth, height=8cm]{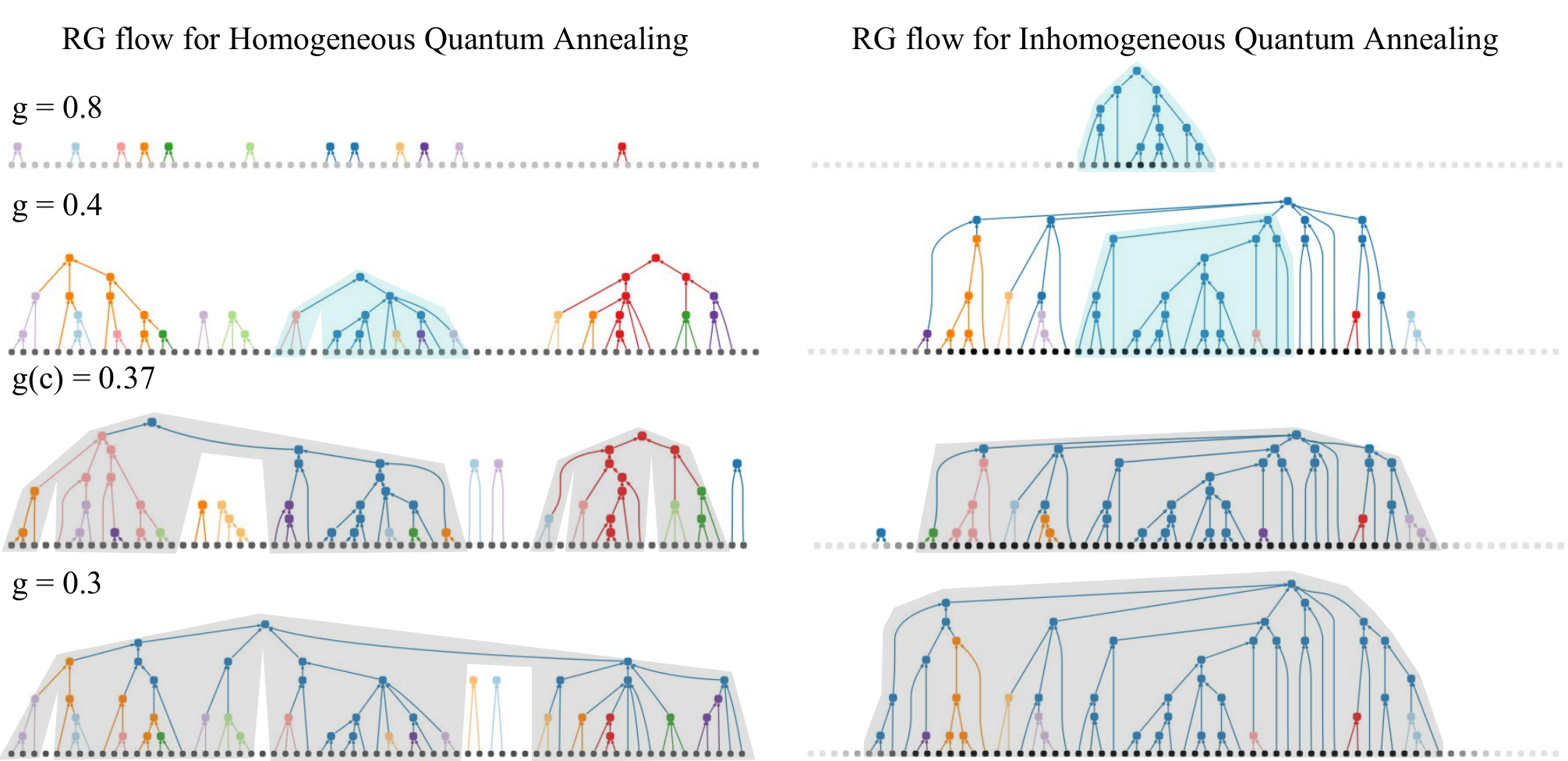}
 \end{center}
  \caption{{ Causal zones formation during RG
      flow:} A tree representation of SDRG cluster formation for four snapshots of
    homogeneous (left) and inhomogeneous (right) quantum annealing at
    transverse fields 
    g = 0.8, 0.4, 0.3 and critical value, $g_{c} =0.37$ for an instance of a 64 spin chain. The RG flow starts 
   with physical spins (as tree leaves represented by dots on a grey
   scale) and ends at the RG fixed point (roots) via other
   macrospins (color dots) in descending energy scales. The black physical spins are those that are
   already in the ordered or symmetry breaking phase while the grey means that they are
   still partially or fully in symmetric phase. The snapshot at g = 0.8 highlight how
   radically different these two strategies are: causally independent cluster formation in HQA and
naturally growing of a central cluster nucleus in IQA. The other
snapshots at $g =0.4, 0.37, 0.3$ are
chosen to be close to the critical point. The shaded blue
area are causal zones that occurs for 0.2 SDRG energy cutoff for
first two panels showing qualitatively and
quantitively different cluster
formations for IQA and HQA. In the last two panels the shaded grey areas
correspond to 
causal zones for the fixed point of SDRG. The key
observation here is that HQA has several causally separated holes in the shaded area where
certain physical spins cannot communicate their choice of symmetry
breaking with other spins. These are the places that
topological defects are highly likely. Nevertheless, within each causal zone
forming at low-energy scale there is also some chance of defects for
macrospins that are causally separated in higher energy scales
(indicated by different colors within each causal zone).}
 \label{fig:sdrg_flow}
\end{figure*}

It should be noted that whenever $z=1$ the threshold velocity becomes constant and we
get a sharp suppression of topological defects whenever we drive the system with critical
front that is slightly below $v_t$. The value of $v_t$ becomes equal to 2 for
a transverse Ising chain when all couplings are equal to one
\cite{Dziarmaga2010a}. This suppression of defects and causal
synchronization, however, could be
affected when we drive the system with a
multi-front strategy.
For a simple pure system driven by
two critical fronts, both moving with velocity smaller than $v_t$, in Fig.~\ref{fig:causalcartoon} we illustrate how the defects can be created when two clusters are
merging together. This highlights the
interplay of the causal effects and different boundary conditions on the defect suppression.

As we have discussed in section \ref{sec:KZM}, the disordered systems have a completely different critical phenomena as $z \nu \to \infty$. In this case one has to modify Eq.~\eqref{eq:iKZM_Gap} accordingly\cite{Rams2016} by taking into account that at the critical point
the gap is expected to vanish as a stretched exponential with the
system size -- effectively given in our case by
Eq.~\eqref{eq:IKZM_xi}. 
We elaborate on this in the next section in the context of multiple critical fronts driving strategy.


\subsection{Universality of causal gaps via SDRG}

In this section we use a combination of analytical and numerical SDRG techniques to show that the distribution of
causal gaps are universal irrespective of the shape of inhomogeneities
for strongly-disorder Ising systems. We also
derive a scaling relation for the dependence of the minimal
causally-induced gap on the actual system size and slope of
inhomogeneity. We first present implementations of SDRG for disordered
spin chains under various schedules. 

The core concept of all RG techniques is to  re-express the
parameters which define a problem by coarse-graining some microscopic
degrees of freedoms. In each step of the RG flow we arrive at 
effective Hamiltonian terms that have fewer and much simpler parameters acting on a
lower energy and larger (macroscopic) length scale, such that certain physical or
computational aspect of interest in the original problem remain
unchanged. The procedure can be recursively repeated until the
Hamiltonian is no longer changing which indicates that we have arrived
at the fixed point of the RG flow.  In SDRG, that is specifically developed for
disordered systems and has been generalized to higer-dimensional systems\cite{Huse_Fisher_2000}, the largest energy
scales is systematically removed via two different operations: site
decimation and bond decimation \cite{Ma_1979,*Dasgupta_1980,Fisher1995,*Fisher1992a}, see \cite{igloi_review_2005} for a review.

A site decimation occurs whenever we have
a site-dependent transverse field which is the largest
energy scale within a local neighborhood of our system; that
is $g_i > J_{ij} \forall j$. Site decimation
means that we basically lock the spin $i$ to the direction of its local transverse
field. Such spin will be effectively decouples from the rest of the
system. Emerging new couplings are generated between all neighbors of the
decimated spin $i$. These effective couplings can be computed within
second order perturbation theory as $ \tilde{J}_{jk} =
\frac{J_{ij}J_{ik}}{g_i}$, if $J_{jk}=0$ or otherwise $ \tilde{J}_{jk} = max(J_{jk},\frac{J_{ij}J_{ik}}{g_i})$.
A bond decimation is performed in similar fashion. A bond decimation
occurs whenever we have two
sites $i$ and $j$ interacting via $J_{ij}$ that is the largest energy
scale within a local neighborhood of our system; i.e. $J_{ij} \geq g_i$
and $J_{ij} \geq g_j$; and also $J_{ij}\geq J_{ik}\forall k$ and $ J_{ij}\geq
J_{lj} \forall l $.  Bond
decimation simply means that we lock the two sites $i$ and $j$ into a
macrospin by projecting the combined pair into their local ground states. No additional bonds or
coupling between any spins are generated in this case. All the spins
that were previously coupled to at least one of  the sites are now interacting with the combined cluster. 
For the spins that were coupled to both $i$ and $j$ we invoke a maximum
selection rule. Effective transverse field 
at the emerging macrospins becomes  $\tilde{g}_{i} =\frac{g_i g_j}{J_{ij}}$.

We apply the above RG procedure to our time dependent Hamiltonian by
considering each time as a snapshot for different instances of static spin
chains, see Fig.~\ref{fig:sdrg_flow}. The
SDRG simulation confirms our assumption of causally independent clusters
in the homogeneous strategy, in the low-energy or long wavelength
limit. In contrast, we observe inter-cluster causal dependence
emerging in the multiple critical fronts strategy, see the SDRG
visualizations of our protocols in Fig.~\ref{fig:sdrg_flow}.  

Next, we discuss an upperbound for a global threshold
velocity such that the multiple critical fronts strategy can lead to suppression of excitation between
the low energy manifold and the rest of excited
states. If we have full parallel annealing in all clusters
simultaneously, we essentially interrupt the causality of symmetry breaking events
between different clusters as illustrated in
Fig~\ref{fig:sdrg_flow}. 
When the fronts are sufficiently separated, the RG flow makes the corresponding clusters exponentially decoupled. Consequently, when looking at the possible transitions, we can consider each front
independently.  The transition matrix elements for each cluster $k$: 
 \begin{equation}
|\langle 0,t | \frac{d
\hat H}{dn_k^f} |1,t \rangle| \frac{dn_k^f}{dt} = \Omega_k(n_k^f) v_k,
\end{equation}
where $n_k^f$ encodes the (time dependent) position of the front and $v_k=\frac{dn_k^f}{dt}$  is its vertical velocity. $\Omega_k(n_k^f)$ defines local instantaneous ground state to first
excited state transition matrix elements for that front.  For simplicity of analysis in
following sections we choose a constant and uniform velocity for all the critical fronts
in various clusters i.e. $v_k=v$.   Due to
causal independence of clusters in multiple criticality, the
adiabatic condition for low-energy states (approximate solutions) 
can be characterized by maximum over of all possible local transition matrix over
its local gap, $\Delta_{k}(n_k^f)$; that is $\frac{\Omega_k(n_k^f)} {\Delta_{k}^2(n^f_k)}$.
Then the low-energy quasi-adiabatic dynamics is expected for 
\begin{equation}
\label{eq:vt_adiabatic}
v \ll v_t =  \min_{k} \frac{\Delta_{k}^2(n^f_k)}{\Omega_k(n^f_k)}.
\end{equation}
In the following we are going to discuss the universality of the
threshold velocity, reflected in its dependence on the shape of
critical front characterized  by the slope $\alpha$.

 \begin{figure} [t]
\begin{center}
  \includegraphics[width=0.95 \columnwidth]{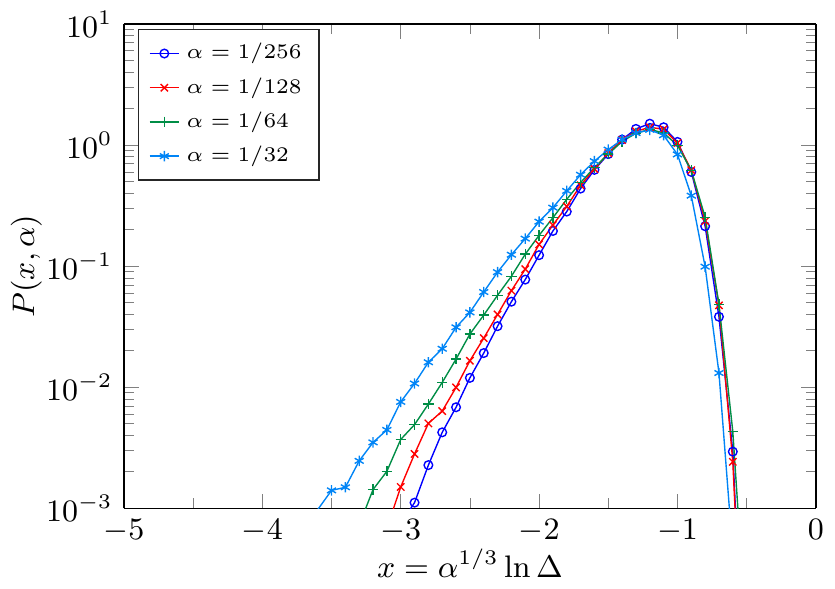}
 \end{center}
\caption{{ Universal scaling of the gap distributions with respect to
 the shape of inhomogeneity.} In order to achieve smooth shapes of
 transverse fields at the border
  of various clusters, we consider
  a type-I inhomogeneity (standing critical wave) with 4
  clusters in a system of $N=1024$. By comparison to Fisher and Young SDRG simulations
  for universality of quantum Ising models \cite{FY1998}, we can
  conclude that our inhomogeneous protocol has changed the effective length scale of system
  from $N$ to $O(\alpha^{-2/3})$. Alternatively, we can consider that
  multiple critical front strategy has changed the critical dynamical
  exponent of full system from $z
\to \infty$ to $z
\to 0$; that is the system always stay close to a critical point
during almost the entire evolution, however it
never experience a true quantum phase transition on the whole
system (shortcutting adiabatic evolution).}
  \label{fig:gapscaling}
\end{figure}

Firstly, it is worth drawing the connection between the condition in Eq.~\eqref{eq:vt_adiabatic} with the threshold velocity which was derived in the previous section.
There, it was calculated as $\hat v = \hat \xi_i / \hat t_i$, where $\hat \xi_i$ and $\hat t_i$ were the characteristic length scale (the penetration depth) and time scales (given by the inverse of the gap) related with the slope of inhomogeneity $\alpha$. Similarly, the adiabatic condition is sometimes formulated as, $v \ll \Delta \cdot \Gamma$, where $\Delta$ is the energy gap, and $\Gamma$ estimates the width of the region (in the driving parameter space)  for which the gap is close to its minimal value, see e.g. \cite{Knysh_2016} -- in analogy to the avoided level crossing and the Landau-Zener problem.  To resolve this seeming inconsistency (i.e. that the gap appearing in Eq.~\eqref{eq:vt_adiabatic} is squared comparing to other expression), we note that $\Omega_k(n^f_k)$ and $\Delta_k(n^f_k)$ are not independent. We expect that $ \Omega_k(n^f_k)/\Delta_k(n^f_k) \sim \xi_i^{-1}$, i.e. it is directly proportional to the size of the effective critical region. We show this in Fig.~\ref{fig:gapscaling} and discuss further below. Importantly, this allows us to focus on the scaling of the gap.

The effective size of the critical region (for a given front) is given by Eq.~\eqref{eq:IKZM_xi}. In our case $\nu=2$, which gives $\xi_i \sim \alpha^{-2/3}$.
The fronts are independent if their respective distances, or the sizes of the clusters, are $\gg \xi_i$. The typical gap related with such critical front is then expected to scale as a stretched exponential in $\xi_i$, $\Delta_k(n^f_k) \sim e^{-\mathrm{const} \sqrt{\xi_i}} \sim e^{-\mathrm{const} \cdot \alpha^{-1/3}}$ \cite{Rams2016}.

To be more precise, we consider $P(\Delta,\alpha)$ as a distribution of the minimal relevant instantaneous gap $\Delta$, as each front is traveling within a cluster for a fixed
$\alpha$. We argue that we can observe the universality of local gap probability distributions by introducing a rescaled log-gap $x =  \alpha^{1/3} \log \Delta$,  with the expected $P(\Delta,\alpha) d\Delta  = \tilde P(x) d x$, and universal distribution $\tilde P(x)$. In contrast, the universal scaling of the gap in the finite size homogeneous system at criticality 
is known to be described by universal distribution $P(\Delta,N) d\Delta  = \tilde P(x) d x $ with $x = N^{-1/2} \log \Delta$ \cite{Young1996,FY1998}. 
Our ansatz is then a directly consequence of an assumption that the effective size of the critical region in our protocol is no longer
given by $N$ but instead is characterized by the penetration depth in Eq.~\eqref{eq:IKZM_xi}.

We verify those scaling predictions in Fig.~\ref{fig:gapscaling}. We consider the system with two clusters, i.e. 4 critical fronts to highlights the independence of the front and that the
scaling prediction naturally carry on to the case of multiple fronts.
We calculate the minimal relevant gap (which would be related with one of the fronts), which we distinguish by finding the minimal energy eigenstate with the corresponding transition matrix element $\Omega$ above some threshold.
We calculate the gap via numerical SDRG which ends at a system of few spins which is subsequently exactly diagonalized. While the SDRG procedure which we use is introducing some errors, we check that the results can be essentially reproduced by the numerically exact solutions based on free-fermionic picture for transverse-field Ising models. 
Using SDRG allows us to highlight that the fronts are independent as they become effectively decoupled during SDRG flow.
We compute the distribution of $\Delta$ by sampling, for each disorder instance,  from different equidistant front positions as they are moving though the chain. We disregard the beginning and the end of the protocol when the system is far from criticality and focus on the relevant intermediate part where we have independent critical fronts. In Fig.~\ref{fig:gapscaling} we collect the statistics using 5000 disorder instances.
Indeed, we can see that the peaks of the rescaled distribution collapse validating the scaling anzats. The tail corresponding to small energies, where the distributions do not properly collapse, is non-universal and results from occasional very weak links (as $J \in [-1,1]$). They give rise to gaps of similar order (occurring again when the magnetic field acting on the neighboring spins is again almost switched off), which are characteristic for strictly $1$D system.  We elaborate more on this later in this section. The presence of such gaps is especially pronounced for larger values of $\alpha$ when the typical gap $\sim e^{-\mathrm{const} \cdot \alpha^{-1/3}}$, which can be attributed to many-body effects, is larger.

 \begin{figure} [t!]
\begin{center}
  \includegraphics[width=0.95 \columnwidth]{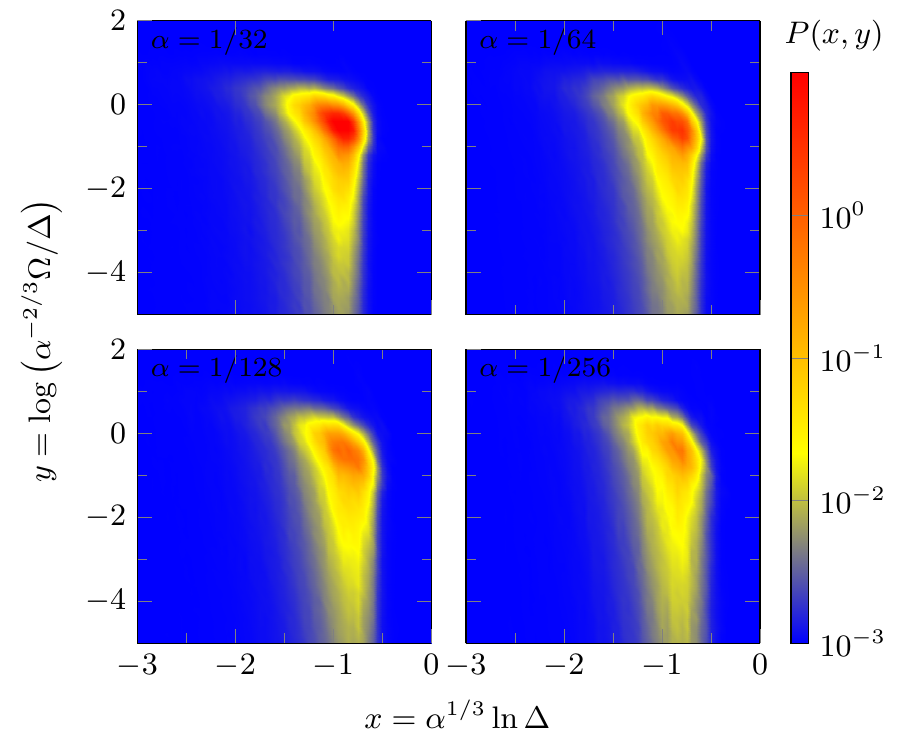}
 \end{center}
 \caption{Inherent dependence of Hamiltonian transition matrix
$\Omega$ and the local gaps $\Delta$, which
can be observed for different inhomogeneity slopes. 
We show the probability density $P(x,y)$, where $x=\alpha^{1/3} \log \Delta$ is the gap recalled gap and $y = \log (\alpha^{-2/3} \Omega/\Delta)$ captures the recalled relation between the mixing term and the gap. The probability densities for different slopes $\alpha$, expressed in the recalled variables $x$ and $y$, are roughly the same. 
We conclude that the maxima of $\Omega/\Delta$ are consistent with the expected scaling of $\Omega/\Delta \sim \xi_i^{-1} \sim \alpha^{2/3}$. }
  \label{fig:omega_delta}
\end{figure}

The analysis above regards local, instantaneous gap related with the critical fronts traveling though the chain.  To quantify the difficulty of the problem, we consider the minimal gap encounter during such quench. To that end we focus on the distribution $P(\Delta,\alpha) d\Delta= \tilde P(x) d x$.
Importantly $\tilde P(x)$, as calculated for the homogeneous critical case in Ref~\onlinecite{Young1996,FY1998}, has a Gaussian tail for $|x| \gg 1$, \footnote{This derivation corresponds to the smallest, single quasi-particle gap in the Ising model. Here due to conserved parity symmetry the relevant gap corresponds to two excited quasi-particle.}  i.e.
\begin{equation}
\tilde P(x) \sim e^{-a x^2}.
\label{eq:prob_gauss}
\end{equation}
Fig.~\ref{fig:gapscaling}, especially in the limit of small $\alpha$ when gaps attributed to weak links are less relevant, indicates that the above holds also in our case\footnote{More precisely we expect the tail vanishing as $\tilde P(x) \sim e^{-a (x-x_0)^2}$, with some non-universal constant $x_0$. We set $x_0=0$ for clarity of derivation as it is not relevant for the main conclusion.}. 
Now, let the probability that $\Delta$ is smaller than some $\Delta_{min}^q$
be 
\begin{equation}
P(\Delta < \Delta^q_{min},\alpha) = \int_{-\infty}^{x_{q}} \tilde P(x) d x= \epsilon,
\end{equation}
where $x_{q} = \log(\Delta^q_{min}) \alpha^{1/3}$. 
In order to find the probability distribution for minimal gap we assume that we
are sampling $N$  times from $P(\Delta,\alpha)$ (or more precisely proportional to $N$ times in the limit of large $N$). That way,
the probability 
\begin{equation}
P(\Delta_{min} > \Delta^q_{min}) = (1-\epsilon)^{N} = q,
\end{equation}
where $\Delta_{min}$ is the minimum from the sample of $N$
instantaneous gaps. This equation defines $\Delta^q_{min}$ as a $q$-quantile
for the global minimal gap. Now, we obtain $\Delta^q_{min}$ from the above equation.
This gives $\frac{1}{N}{\log q} = \log(1-\varepsilon) \simeq -\epsilon \sim \frac{1}{2a x_q} e^{-a x_q^2}$, which is obtained by expanding the error function for the Gaussian tail in Eq.~\eqref{eq:prob_gauss}, to the
leading order in small $x_q$. Solving this equation in the leading
order we obtain 
\begin{equation}
\alpha^{1/3} \log \Delta^q_{\min} = x_q \approx -\sqrt{ \frac{1}{2a} \log \frac{N} {2a \log q^{-1}} }. 
\label{eq:Delta_min_q}
\end{equation}
This suggests that if we fix a quantile $q$, then in the asymptotic
limit
\begin{equation}
\Delta^q_{min} \sim \exp(-\mathrm{const} \cdot \alpha^{-1/3} \cdot \sqrt{\log N}), 
\label{eq:gapscaling}
\end{equation}
i.e. it is vanishing slower than any polynomial with increasing
$N$. This has to be compared with homogeneous gap scaling as:
$\Delta^q_{min} \sim \exp(-\mathrm{const} \cdot \sqrt{N})$, which vanishes as stretched exponential with $N$. We note here that similar
analysis in case when $\tilde P(x)$ would have exponential tail for
large negative $x$ would give polynomial dependence for the minimal
gap on $N$.

In order to fully analyse the tail of the gap distribution we also
consider the situation that the minimum gap is enforced by very small local
link rather than many-body gap of critical system of effective size $\xi_i$. This issue, which is
essentially an artifact of 1D systems, can be largely
avoided by a multi-front strategy where such links are placed
in-between clusters which are driven quasi-adiabatically.
To that end, let's assume that the links are drawn from uniform distribution $J_i\in[-1,1]$.
Probability that a single link is weaker then some $\epsilon$ is equal $\epsilon$, or
$P(|J_i| > \epsilon) = 1- \epsilon$. Let's consider that $q$ is the
probability that all the links are stronger than this $\epsilon$ is $q
= (1-\epsilon)^N$. For small $\epsilon$ this yields
$$
\epsilon \simeq -\frac{\log q}{N}.
$$
The minimal gap related with such a weak link is of similar size, and such effects become relevant in $1$D geometry.
Finally, we should note that if we consider the distribution of the logarithm of the gap $x$, the uniform distribution and related weak links directly translate to the exponential tail of $P(x)$ -- mentioned in the previous paragraph -- which is indeed still visible in Fig.~\ref{fig:gapscaling} for larger values of $\alpha$.

 Finally, in Fig.~\ref{fig:omega_delta} we illustrate the relation between the  transition matrix elements $\Omega$ and the gap $\Delta$. To that end, for simplicity we consider protocol with single front
 and, as in Fig.~\ref{fig:gapscaling}, sample the values of minimal relevant gap and corresponding $\Omega$ as the front is traveling though the chain. The results are collected as probability distribution, where the statistics is collected from 5000 instances. In order to illustrate the universal behavior we employ rescaled variables. For the gap $x=\alpha^{1/3} \log(\Delta)$ as above, and $y = \log(\alpha^{-2/3} \Omega/\Delta  )$ to reflect the expected relation $\Omega/\Delta \sim \xi_i^{-1}\sim \alpha^{2/3}$.  We plot the obtained distributions of $P(x,y)$ in Fig.~\ref{fig:omega_delta} for several different values of $\alpha$. We observe that they are roughly similar in agreement with our prediction.
 We should note that local maxima of $\Omega/\Delta$ -- i.e.  where it is most relevant -- coincide with the local minima of energy gap. Apart from those point, $\Omega/\Delta$ is quickly approaching zero, which is reflected by elongated shape of the distribution $P(x,y)$ in the direction of small (irrelevant) $y$.


\section{Random cluster-Ising Hamiltonian}
\label{sec:clusterH}

The numerical results in the previous sections were confined to 1D geometry where, in the final ground state, the neighboring
spins are aligned according to corresponding $J_{n,n+1}$.  
In this section we show that the general approach discussed in this
article does 
not hinge on the possibility to align nearest-neighbor interacting
spins in strictly 1D Ising geometry.  To that end we consider random
cluster-Ising Hamiltonian as follows:
\begin{eqnarray}
H(t) &=& - \sum_{n=1}^{N-1} J_n  \sigma^z_n \sigma^z_{n+1}
         -\sum_{n=1}^{N-2} K_n   \sigma^z_n \sigma^x_{n+1}
         \sigma^z_{n+2} \nonumber \\ 
 &&- \sum_{n=1}^{N} g(n,t) \sigma^x_n, 
\label{eq:k-local}
\end{eqnarray}
where the first two terms contains the problem Hamiltonian and the
last term is the (inhomogeneous) driving term given by external transverse field. 
The quench dynamics generated by this Hamiltonian can be simulated analogously as for the Ising chain from the previous section
as -- using Jordan-Wigner transformation -- it can be mapped onto a
free fermionic system.

\begin{figure} [t]
\begin{center}
  \includegraphics[width=0.95 \columnwidth]{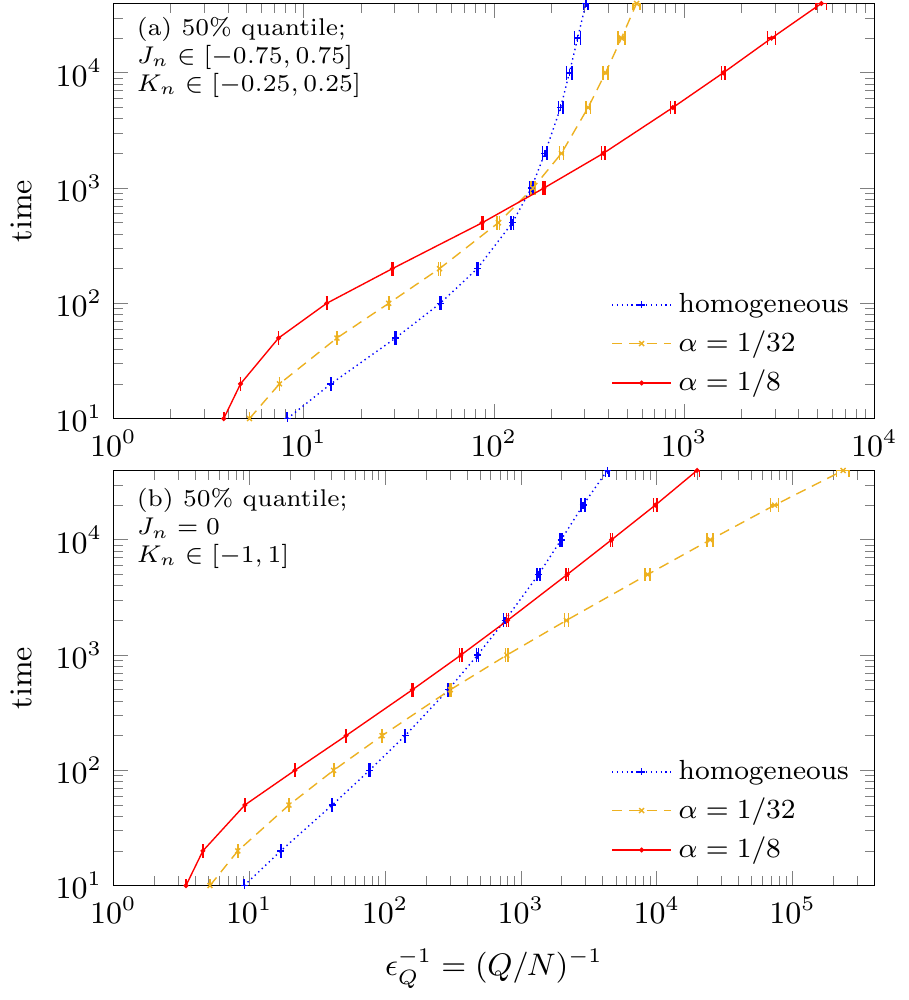}
 \end{center}
  \caption{{ Scaling of the total annealing time as a function of inverse
      residual energy for homogeneous control (blue lines) versus
    inhomogeneous protocol:} Here we use two fronts spreading from the
  center of the chain for $N=256$. The Ising Hamiltonian in
  Eq.~\eqref{eq:k-local} has generally three different paramagnetic,
  ferromagnetic and topological insulator phases. In (a) the
  Hamiltonian contains
  2- and 3- local random terms versus in (b) only 3- local random 
  terms are present. It can be observed that
in both cases non-zero $\alpha$ significantly outperform the
homogeneous or $\alpha=0$ schedule, with an exponential advantage of
optimal inhomogeneous protocol $\alpha=1/8$
when we have an interplay of two- and three-body interactions.}
 \label{fig:XZX}
\end{figure}

First, lets consider the pure and homogeneous model with $J_n = J>0$, $K_n = K>0$ and $g_n = g>0$.
Depending on the relative strength of those terms the model has 3 distinct phases, see e.g. \cite{Song_2015,Titvinidze2003,verdal2011a,verdal2011b,Suzuki_1979}. It is convenient to set $K+J=1$.
When the magnetic field is dominating, $g > 1$, the system is in paramagnetic phase analogously to the Ising model. We are going to initialize the evolution in this phase. In the opposite limit, $g<1$, the system is in ferromagnetic phase for $J+g > K$ and symmetry-protected topological order phase for $K>J+g$. For $J=g=0$ it reduces to the cluster Hamiltonian \cite{Briegel_2001}.
The phases are separated by critical points with $z=1$ or $z=2$.

We consider random couplings $J_n$ and $K_n$, which makes the final
target state far from being trivial for a 256 spin system. 
We present the results of the quench in Fig.~\ref{fig:XZX}. We consider two different disorder distributions: 
in Fig.~\ref{fig:XZX}(a) $J_n$ are dominating -- that is $J_n \in [-0.75,0.75]$  and  $K_n
\in [-0.25,0.25]$, and in Fig.~\ref{fig:XZX}(b) all $J_n=0$ and $K_n
\in [-1,1]$. For homogeneous driving the residual energy is vanishing
logarithmically with the quench time in both cases, which is a similar behavior as for the random Ising model. 
It can be observed that there is crossover of the performance for
sufficiently long times and the homogeneous protocol is considerably outperformed by 
an inhomogeneous driving fields with the optimal slope. The advantage
for the case of random $J_n$ could be exponential, see
Fig.~\ref{fig:XZX}(a). In this case, the spatial inhomogeneity allows
the system to reach the quality of solution (small residual energies)
which are practically unattainable within the homogeneous
approach. Here, we use a version with single cluster and two critical
fronts spreading from the center of the chain.  It should be noted
that the optimal shapes ($\alpha$) in those two cases are
different. This optimal value of $\alpha$ can be found numerically for given
distribution of disorders as a simple preprocessing, or hyper-parameter characterization,
similar to the spirit of finding the optimal annealing time or number
of sweeps for simulated annealing or quantum Monte Carlo
solvers.


\section{Conclusion and future works}
\label{sec:conclusion}

We have presented a model for engineering quantum phase transitions in
disordered systems by manipulating information flow among
clusters that are formed within a quantum critical region. We have shown that
space-time inhomogeneities in the control fields could lead to
reconstruction of causal zones (light cones), such that symmetry
breaking events can be synchronized suppressing the density of topological
defects and/or redistributing their spatial arrangements. We have used exact diagonalization
techniques for 1D systems to show an exponential speedup of
non-adiabatic inhomogeneous quantum annealing over standard adiabatic quantum
computing, even in the presence of higher order interactions. By application of
renormalization group techniques we have demonstrated that the effective
causal gaps exhibit universality with respect to the shape of inhomogeneity. We have derived a scaling relation showing such
effective gaps have sub-polynomial scaling with
the system size, in contrast to stretch exponential for homogeneous
control strategies.  In a subsequent
work \cite{Mohseni2018}, we will provide a detailed theoretical
discussion of our work as a generalization of KZM for disordered systems including various bounds for
the shape of critical fronts and threshold velocities under different
assumptions.  We will also discuss how our approach can be applied to
low-dimensional spin-glass problem Hamiltonians 
\cite{MohseniAQC2017,Mohseni2018}. During the preparation of this manuscript an
exponential speedup for inhomogeneous quantum annealing of p-spin model 
was reported showing ferromagnetic first-order phase transitions
can be smeared with inhomogeneous control strategies \cite{Nishimori2018}.

\textit{Acknowledgement.} We would like to thank Sergio Boixo, Adolfo
del Campo, Hartmut Neven, Susanne Pielawa, and Vadim Smelyanskiy for useful discussions.
M.M.R. acknowledges support by National Science Center Poland under Projects No. 2016/23/D/ST3/00384,
as well as receiving Google Faculty Research Award 2017.
M.M.R acknowledges using the  supercomputer  ``Deszno''  purchased  thanks  to  the  financial  support  of  the  European  Regional  Development  Fund  in
the  framework  of  the  Polish  Innovation  Economy  Operational   Program   (contract   no.   POIG.   02.01.00-12-023/08).



%

\end{document}